# On the Posch ratio for irradiance in coastal waters and the high seas

Xabier Pérez-Couto,[1,*] Fabio Falchi,[2,3] and Salvador Bará[1]

[1] *Agrupación Astronómica "Ío", 15005 A Coruña, Galicia*

[2] *Photonics4Life, Departamento de Física Aplicada, Universidade de Santiago de Compostela, 15782 Santiago de Compostela, Galicia, Spain*

[3] *Istituto di Scienza e Tecnologia dell'Inquinamento Luminoso (ISTIL), 36016 Thiene, Italy*

[*] *e-mail: xabierperezcouto@gmail.com*

**ABSTRACT**

The horizontal irradiance at the sea surface is an informative light pollution indicator to study Artificial Light at Night (ALAN) effects on marine biodiversity (e.g.: zooplankton diel vertical migration). The Posch ratio (PR) for the horizontal irradiance (that is, the ratio of the horizontal irradiance to the zenith radiance) is a useful tool for estimating the irradiance from easily available measurements of the zenith night sky brightness. The PR definition has already been generalized for any pair of linear radiance indicators in any pair of arbitrarily chosen photometric bands, and can also be applied to estimate e.g. the average sky radiance or the radiance at some elevation above the horizon as a function of the radiance in any other direction of the sky. The PR for a single light source depends on the distance from the source, its angular and spectral emission pattern, and the state of the atmosphere. The PR for any set of sources is a linear combination of the individual PRs that each one would produce separately, with weights that can be easily derived from the relative contribution of each source to the zenith radiance. Whereas in populated lands the ALAN PR varies relatively fast from one location to another, due to the particular spatial distribution of lights, in coastal waters and the high seas the light pollution PR is a smooth function of the distance to the shoreline, due to the progressive lack of neighboring sources and the absence of obstacles. In this work we present the fundamental equations of the model and an example of application for the waters surrounding the Iberian Peninsula, North Africa and the West Mediterranean islands.

ORCID

X. Pérez-Couto https://orcid.org/0000-0001-5797-252X

F. Falchi https://orcid.org/0000-0002-3706-5639

S. Bará https://orcid.org/0000-0003-1274-8043

**1. INTRODUCTION**

The continued increase of artificial light at night (ALAN) in radiance and extension each year across the globe [1] in combination with the present and future growth of population in coastal areas [2] has accentuated the presence of ALAN both in coastal waters and high seas, due to the the emissions of light by nearshore towns, cities as well as its harbors [3], either by the direct illumination of water masses or by the scattering of light in the atmosphere and its propagation back to the sea surface [4] reaching tens or even hundreds of kilometers out to sea [5].

At the same time, the influence of ALAN in marine ecosystems has been a topic of concern in ecological and environmental studies for decades [6-9], as it has been shown to affect to predator-prey systems of estuary fishes [10], interactions between intertidal invertebrates [11], sea turtle nesting [12], birds strikes with ships [3], and zooplankton diel vertical migrations [13, 14], probably the biggest daily migration of biomass on the planet [15].

Among all the light pollution indicators defined and applied in the literature to analyze the impact of ALAN in marine ecosystems, the sea surface horizontal irradiance, i.e. the amount of power arriving to the sea surface per unit surface [16], is one of the most informative in order to study the reach and impact of biologically important ALAN in coastal waters and high seas, both in the surface as in depth, as seen in Davies et al. (2020) [17] and most recently in the Global Atlas of ALAN under the sea by Smyth et al. (2021) [18]. There are also some works which address successfully measurements of light pollution indicators from a boat, e.g. in Jechow et al. (2017) [19] all-sky fisheye photometry was used so they could obtain several linear radiance indicators, and in Ges et al. (2018) [20] an improved gimbal system was developed for zenith sky brightness measurements from a moving and unstable boat.

With the aim to obtain new insights about how to estimate the sea horizontal irradiance due to artificial light from more easily available measurements such as the zenith sky brightness, we carried out a study of the behaviour of the Posch Ratio in the sea. The Posch Ratio (PR), named after deeply missed Thomas Posch (1974-2019), is defined in general terms as the ratio of any two linear radiance indicators [21] and has been studied recently in the literature: first by Kocifaj et al. (2015) [22] for three canonical atmospheric cases, later by Jechow et al. (2020) [23] and Davies et al. (2020) [17] with some approximations based in observational data and, lately, by Bará et al. (2022) [21] where its algebraic properties were described as well as its behaviour in natural and in-land light polluted skies, been proved as a useful tool in emerged lands providing good estimations of the horizontal irradiance and the average all-sky radiance as a function of the zenith sky brightness.

Taking advantages of some characteristic features of the marine environment with respect to the terrestrial such as are the absence of light sources in the space located between the observer and the coastline and the non-existence of obstacles, we present the equations of a simplified sea-coastline model of the PR in section 2. Subsequently, some maps encompassing the waters surrounding the Iberian Peninsula, North Africa and West Mediterranean islands were produced in order to perform several statistical analysis aiming to get some practical insights about the Posch Ratio distribution and application in a real sea environment.

## 2. A SIMPLIFIED SEA-COASTLINE PR MODEL

Let $B_K(\mathbf{r})$ be any radiance-derived indicator (e.g. zenith sky brightness or horizontal irradiance) at the observer position $\mathbf{r}$, it can be obtained from the emissions $L_1(\mathbf{r}')$ of the sources located at points $\mathbf{r}'$ within the territory $A'$ as [21]

$$B_K(\mathbf{r}) = \int_{A'} K(\mathbf{r},\mathbf{r}') \, L_1(\mathbf{r}') \, d^2\mathbf{r}' \qquad (1)$$

where $d^2\mathbf{r}'$ is the element of surface (in m²) and $K(\mathbf{r},\mathbf{r}')$ is the point spread function (PSF) for the corresponding indicator. The dimensions of $K(\mathbf{r},\mathbf{r}')$ may be different for different indicators and different input information about the sources $L_1(\mathbf{r}')$, and are given by $[K] = [B]/(m^2 \cdot [L_1])$. A general Posch ratio has then the form

$$\Pi_{K0} = \frac{B_K(\mathbf{r})}{L_0(\mathbf{r})} = \frac{\int_{A'} K(\mathbf{r},\mathbf{r}') \, L_1(\mathbf{r}') \, d^2\mathbf{r}'}{\int_{A'} K_0(\mathbf{r},\mathbf{r}') \, L_1(\mathbf{r}') \, d^2\mathbf{r}'} \qquad (2)$$

where $L_0(\mathbf{r})$ is the zenith sky radiance and $K_0(\mathbf{r},\mathbf{r}')$ the PSF of its propagation.

Due to the absence of obstacles and light sources for a marine observer between the shoreline and their position, a simplified model of the marine-coastal environment can be obtained by considering a straight coastline with sources located in-land. An even simpler but still not too unrealistic model is the one with uniform spatial density of sources within a strip of territory of certain width $W$ (see figure 1). The uniform source distribution has to be understood here in a local average sense, i.e. even if the sources are points of light distributed in a very non uniform way at the microscale (e.g. tens of meters), they average to approximately the same amount of emissions when considered at larger scales that are nevertheless still small enough as to be identified with the terms $d^2\mathbf{r}'$ in the equations above (e.g. the pixels of the VIIRS when the observer is a few kilometers away from them).

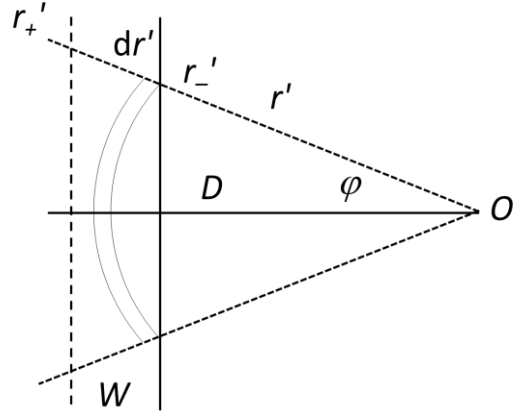

**Figure 1.** Scheme of the simplified model where the shoreline is straight and in-land light sources are uniformly distributed within a strip of width W while the observer O is located at a distance to the shoreline D.

The calculation of the indicators is highly simplified in this case, especially with rotationally symmetric Point Spread Functions (PSFs), that characterize how each indicator varies in function of the distance of an observer to a single source; because their values are given by expressions like:

$$B_K(D) = L_1 \int_{\varphi_-=-\frac{\pi}{2}}^{\varphi_+=+\frac{\pi}{2}} \int_{r'_-=\frac{D}{\cos\varphi}}^{r'_+=\frac{D+W}{\cos\varphi}} K(r')\, r'\, dr'\, d\varphi \qquad (3)$$

where the origin of coordinates is located in the observer position (so $\mathbf{r} = 0$ and $|\mathbf{r} - \mathbf{r}'| = r'$), $D$ is the distance (for this case we count it in km) of the observer $O$ to the coast and $W$ the strip width. In this work the PSFs are calculated using the hemispheric radiance distribution computed in 100 000 sky points observed at several distances from a single light source with the Garstang–Cinzano atmospheric model, and then angularly integrating the resulting hemispherical data points (in Zenith Equal Area projection) with different weights depending of the indicator to obtain the PSF value for each distance and for three indicators (zenith sky brightness, horizontal irradiance and average all-sky radiance) as described with full detail in Falchi & Bará (2021) [24]. Then, assuming that both sources and the observer are located at sea level (reasonable for many coastal areas) we get a good fit for the Johnson-Cousins V band with an atmospheric clarity parameter K=1 (which represents, excepting for a numerical scaling constant, the ratio of the

aerosol and molecular total cross sections, see eq. 4 in [25], meaning, in the case of K=1, that we are considering a clear atmosphere at sea level. It should not be confused with the symbol for the PSF functions, $K$) of log PSF versus log distance $r'$ with a polynomial of 6$^{th}$ degree for each indicator. Therefore, the PSF can be expressed by means of polynomial powers of ten (polynomials in powers of $\log_{10}(r')$ of the type:

$$K(r') = 10^{\log_{10} K(r')} = 10^{\sum_{k=0}^{n} a_k (\log_{10} r')^k} \quad (4)$$

typically of order *n*=6 or so.)

Note that when taking the ratio of two indicators the factor $L_1$ cancels out and the Posch ratios in this highly basic model are simply evaluated from the PSFs.

The fitting coefficients (from exponent 6 to 0) for the polynomials of 6th degree in log r (being r in km) corresponding to the three PSFs (one per indicator) are shown in Table 1:

| Indicator | log-log polynomial fit coefficients $(a_n \ldots a_0)$ for $(\log_{10} r')^n \ldots (\log_{10} r')^0$ | | | | | | |
|---|---|---|---|---|---|---|---|
| $(\log_{10} r')$ *power* | 6 | 5 | 4 | 3 | 2 | 1 | 0 |
| 'AvAllSky' | -0.0558 | 0.2720 | -0.2786 | -0.4142 | 0.0562 | -1.1363 | 0.1695 |
| 'ZSB' | -0.0394 | 0.1673 | -0.1622 | -0.2202 | -0.0388 | -1.4261 | -0.4436 |
| 'HorIrr' | -0.0424 | 0.1915 | -0.1831 | -0.2996 | -0.0377 | -1.2493 | 0.3993 |

**Table 1.** Log-log polynomial fit coefficients (from exponent 6 to exponent 0) of the PSFs of three linear radiance indicators (Average All-Sky Radiance, Zenith Sky Brightness and Horizontal Irradiance) with r in km.

The numerical integrals of the horizontal irradiance and average all-sky radiance Posch ratios with the above PSFs were performed until a distance to the shoreline of *D*=300 km and for coastal strips of width *W*=1 km, 10 km and 100 km as is seen below (Figure 2).

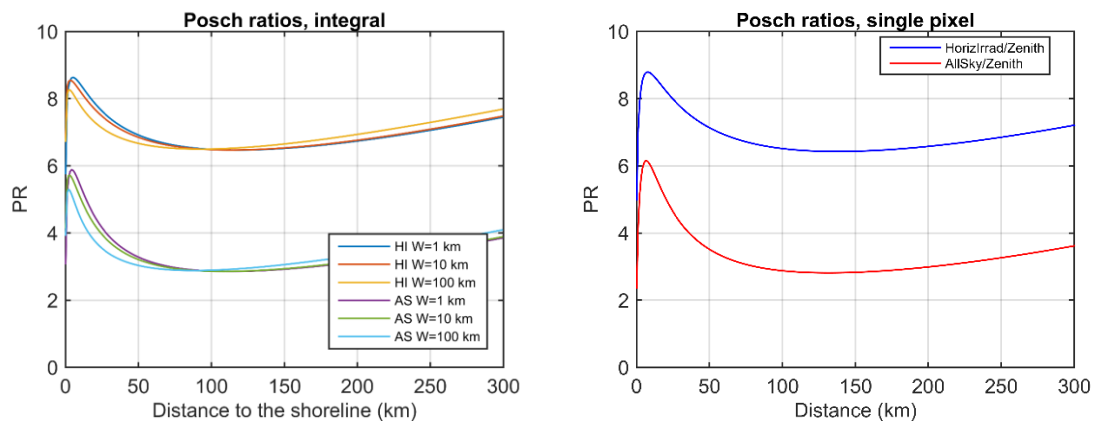

**Figure 2.** Sea horizontal irradiance (HI) and average all-sky radiance (AS) Posch ratios (in sr and dimensionless respectively) versus distance to the shoreline (in km) for several width strips (W=1km, 10km, 100km) (left) and sea horizontal irradiance and average all-sky radiance Posch

ratios (in sr and dimensionless respectively) versus distance to the shoreline (in km) for a single point source (right).

As is seen in Figure 2 (left) the PRs experiences an abrupt and rapid variation in the first 50 km from the coast, especially in the very first kilometers, where the PRs increase very steeply until reaching their maximum, for subsequently decrease and begin to stabilize within a shorter range of values. This is determined by the behavior of the Posch ratio for point sources as a function of the distance to the observer. As is shown in Bará et al. (2022) [21] the overall Posch ratio is due to the weighted sum of the Posch ratios produced by each source separately. For a given emitting pixel of finite area $\Delta^2 \mathbf{r}' = r' \Delta r' \Delta \varphi$ the Posch ratio turns out to be:

$$\Pi_{K0, single\ pixel}(r') = \frac{B_K(r')}{L_0(r')} = \frac{K(\mathbf{r}, \mathbf{r}')\, L_1(r')\, \Delta^2 \mathbf{r}'}{K_0(\mathbf{r}, \mathbf{r}')\, L_1(r')\, \Delta^2 \mathbf{r}'} = \frac{K(r')}{K_0(r')} \tag{5}$$

and, as shown in Figure 2 (right), has a similar behaviour as the previous numerical integrals.

## 3. APPLICATION TO AN ACTUAL COASTLINE

In order to perform some statistics to understand the distribution of the expected marine Posch ratios in sea environments with real and not necessarily straight coastlines, several radiance-derived indicators maps (horizontal irradiance, average all-sky radiance and zenith sky brightness) were produced encompassing the waters surrounding the Iberian Peninsula, North Africa and the West Mediterranean islands (Figure 3). All maps were calculated using the results shown in Falchi & Bará (2021) [26] with the shift-invariant, rotationally symmetric PSFs described in the previous section (Table 1) in combination with VIIRS night-time lights imagery data, free of clouds, moonlit, outliers, stray lights and ephemerical lights, and averaged for the year 2015 [27]; and with Fast Fourier-transform techniques aiming to perform efficient and low-time consuming calculations of the output rasters, as is fully detailed in Bará et al. (2020) [26].

As shown in the density histograms (Figure 4, top row) of horizontal irradiance ('HorIrr') and average all-sky radiance ('AvAllSky') Posch ratios, there is a favourite mean PR in each case: 6.87 sr (sd = 4.98e-01) as 'HorIrr' PR and 3.33 (dimensionless, with sd = 6.48e-01) in the 'AvAllSky' PR case, with less probable values as they increment. At the same time the sea HorIrr PR lies within a range of [5.67, 8.71] (sr) and the AvAllSky PR in [2.84, 6.03] (dimensionless).

Please note that the standard deviation (sd) is calculated as follows:

$$sd = \sqrt{\frac{\sum_{i=1}^{N}(x_i - \bar{x})^2}{N-1}} \tag{6}$$

where $x_i$ is the Posch Ratio for one of the selected indicators in the pixel $i$, $\bar{x}$ is the mean PR over all pixels and $N$ represents the sample's size.

This mean values in combination with the distribution seen in Figure 4 (bottom row) of the PRs in function of the zenith sky brightness is consistent with the fact that the smaller and mean PRs are concentrated, above all, in the high seas, meanwhile the higher PRs correspond to a much wider range, locating from high seas to coastal areas, as the numerical integrals of the model computed in section 2 anticipated.

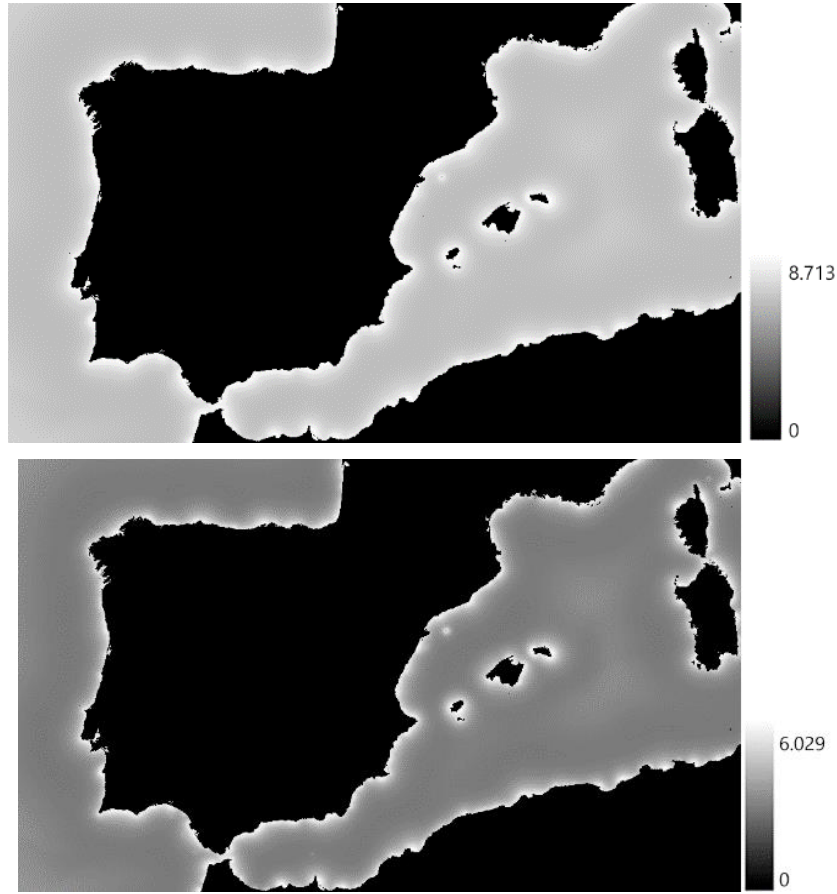

**Figure 3.** Posch ratios for the sea horizontal irradiance in (in sr, top) and for the average hemispheric radiance (dimensionless, bottom) in the Johnson V band for the waters surrounding the Iberian Peninsula, North Africa and the West Mediterranean islands.

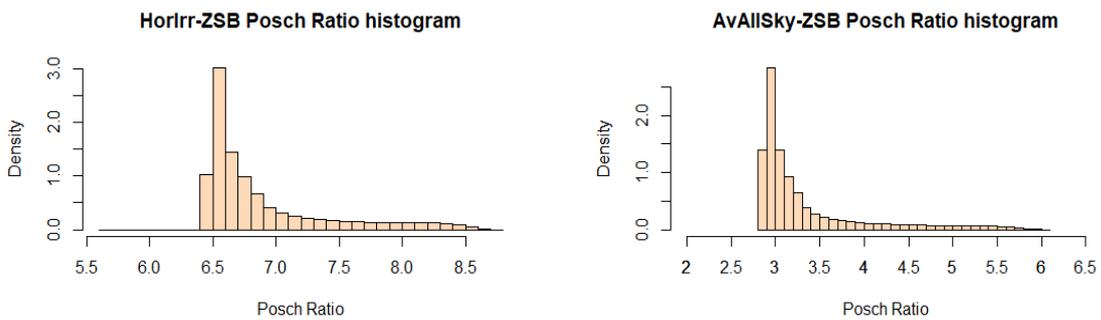

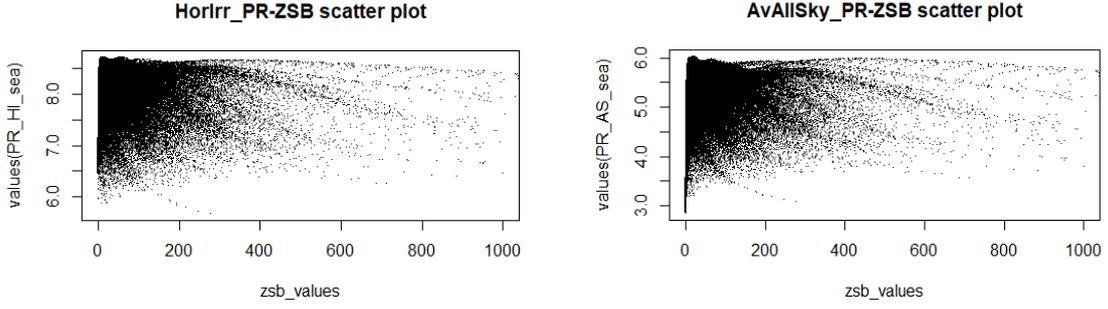

**Figure 4.** Top row: Sea horizontal irradiance to zenith sky radiance Posch ratio (in sr, left) and average all-sky radiance to zenith sky radiance Posch ratio (dimensionless, right) density histograms. Bottom row: Sea horizontal irradiance Posch ratio (sr) to zenith sky brightness (arbitrary linear units, left) and average all-sky radiance Posch ratio (dimensionless) to zenith sky brightness (arbitrary linear units, right) scatter plots.

In order to test numerically the validaty and accuracy of the straight coastline approximation we produced a raster of distances to the atlantic Galicia and Portugal shoreline up to 190 km in QGIS v.3.20.3. Then, the numerical integrals described in Section 2 were computed aiming to obtain the values of the Posch ratio for sea horizontal irradiance each 1km from the nearest point of the coast (width coastal strip of $W$=10km), producing a raster of the galician-portuguese waters where each pixel has the value of the Posch Ratio corresponding to its distance to the coastline.

Subsequently, this raster was multiplied by the in-extent-equivalent zenith sky brightness raster computed to obtain the Posch ratio maps showed in Figure 4, getting a prediction of the values of the sea horizontal irradiances (see Figure 5). Once compared with the one used to calculate the mentioned PR map the mean relative estimation error over $N'$=1 301 111 pixels, computed as

$$\frac{\sum_{i=1}^{N'}\left|\frac{\widetilde{HI}_i - \widehat{HI}_i}{\widehat{HI}_i}\right|}{N'} \tag{7}$$

being $\widehat{HI}_i$ the sea surface horizontal irradiance calculated for the $i$-pixel considering the real coastline geometry and VIIRS-DNB data, and $\widetilde{HI}_i$ the one predicted for the same pixel with our straight shoreline and coastal uniformly distributed light sources approximation.

The obtained mean relative error was 0.019 (sd=0.021). The same operation was performed to estimate the average all-sky radiance, resulting in a mean relative estimation error of 0.045 with sd=0.052.

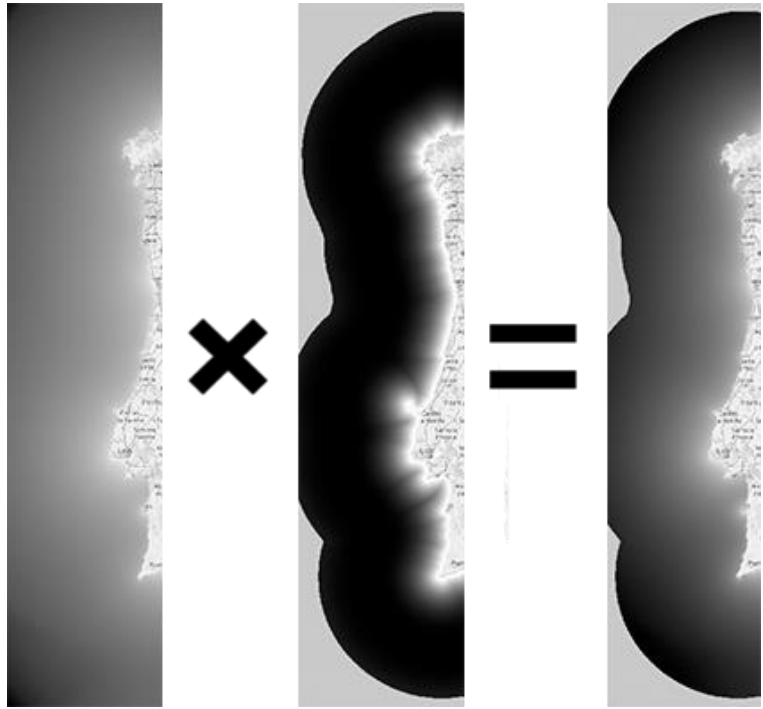

**Figure 5.** Zenith sky brightness raster of the galician-portuguese seas in arbitrary consistent linear units (left), Posch ratio raster for the sea horizontal irradiance (in sr) up to 190km from the shoreline computed with $W$=10km (center) and the resulting sea horizontal irradiance estimation raster in arbitrary consistent linear units (right). In the three plots what we are actually representing is the $\log_{10}$ of each raster in order to facilitate the visualization of the area of interest (the galician-portuguese seas) but the multiplication shown has to be applied on the non logarithmic rasters.

## 4. DISCUSSION

The resulting numerical integrals computed from the sea-coastline model presented in section 2 shows that the artificial component of the Posch ratio undergoes a steep rise and fall in the first kilometers from the coastline, varying in around two units of magnitude at the first 50km, for later begins to behave in a softer way. This suggest that take a mean value as PR to estimate any linear radiance indicator from another in high seas may serve as a first approximation, but in coastal waters definitely not.

This have to be taken into account when a regression analysis is attempted over the sea surface light field. In [18] the sea surface irradiance used to compute the light pollution levels at several depths through the water column were predicted with slopes from a quantile regression calculated in [17], using irradiance data from a measurement campaign across Plymouth Sound and the Tamar Estuary coastal waters, in addition to the predicted zenith night sky brightness from the

New World Atlas [5]. While we agree that a quantile regression is a useful method to remove outliers, the fitted line only will give a realistic outcome when the variables would be measured in the same atmospheric conditions and within the same spectral band. In addition to this, as the results in this work show, the fitted model would be only representative of the irradiance in coastal waters at the same distance to the shoreline as in the sample surveyed in [17] cannot being generalized to the whole coastal area.

Furthermore, the choice of the coastal strip width has an important weight in the range of values in which the PR moves, reaching a smaller maximum value when considering larger coastal strips, being consistent with the fact that the zenith sky brightness (the Posch ratio's denominator) should be a little higher due to the artificial skyglow scattering component of the most distant light sources.

On the other hand, the marine PR calculated from the FFT maps were in the range [5.67, 8.71] for horizontal irradiance and [2.84, 6.03] for average all-sky radiance, clearly differing of the ones calculated for emerged lands in Bará et al. (2022) [21] where the PRs lies within [3.65, 7.59] and [1.36, 4.64] respectively, showing as well as a different frequency distribution as could be seen comparing the corresponding histograms in figure 4 (top). This again justifies the need to study the marine environment as a separated case from the terrestrial one. In fact, the small mean relative estimation errors obtained implementing the sea-coastline model in the galician-portuguese seas shows that it is worth considering the specific characteristics of the environment we want to study, in this case the sea, but being able to apply to some in-land areas. For example, applying different approaches to urban nucleis and to large natural parks and dark protected areas where, due to the absence of direct light sources, a model based on the one studied in this work, just adding an obstacle component, may be useful.

In addition to this and with the purpose of having, not only better estimations but also a deeper knowledge about the mutual behaviour that can exists between each pair of radiance-derived indicators, the next steps should be on the line of a) study the influence of different atmospherical configurations in the PR (e.g. different AOD, aerosol composition, albedo and scattering phase functions…) b) expand the broadband PR to a diverse variety of spectra with the aim of be more practical to ecological and environmental studies and c) support this theory with observational data from diverse territories, for instance through research campaigns like those made in [19] and [20], which would measures the PR empirically in order to compare it with the expected values.

## 5. CONCLUSION

The results in this work show that considering a straight coastline together with uniformly distributed light sources within a strip of coast with a certain width to compute some linear radiance indicators (in this work the horizontal irradiance and the average all-sky radiance) at certain distances from the shoreline and then use their Posch Ratios to the zenith sky brightness to estimate the horizontal irradiance and the average all-sky radiance from the zenithal one, we can get values approximate to within a ~5% (in average) of those obtained if we perform the calculations considering the real coastline geometry and the actual light sources distribution from VIIRS-DNB data.

Therefore, and notwithstanding that making a linear regression may be good enough to have a first approximation to estimate the horizontal irradiance from the zenith sky brightness, a quadratic one or, even better, a model which takes into consideration the physical distribution of the PR based on the features of the area of interest, as the simplified model for marine PR we provide in this work does, seems to be more realistic and accurate.

## ACKNOWLEDGEMENTS

This work has been partially supported by Xunta de Galicia/FEDER, grant ED431B 2020/29.